\documentclass[article]{elsarticle}

\newcommand{\virg}[1]{``#1''}   
\newcommand{\ket}[1]{\left\vert#1\right\rangle}
\newcommand{\bra}[1]{\left\langle#1\right\vert}

\newcommand{\opkb}[2]{\left\vert#1\right\rangle\!\!\left\langle#2\right\vert}
\newcommand{\Trace}[1]{Tr\left[#1\right]}

\usepackage{bbold}
\usepackage{amsmath}
\usepackage[super]{nth}
\usepackage{pdfpages}
\usepackage{xcolor}

\journal{Journal of \LaTeX\ Templates}

\begin{document}

\begin{frontmatter}

\title{Multipartite entanglement transfer in spin chains}

\author[add1]{Tony J. G. Apollaro}
\ead{tony.apollaro@um.edu.mt}
\author[add1]{Claudio Sanavio}
\author[add2]{Wayne Jordan Chetcuti}
\author[add3,add1]{Salvatore Lorenzo}
\address[add1]{Department of Physics, Faculty of Science, University of Malta, Msida MSD 2080, Malta}
\address[add2]{Dipartimento di Fisica e Astronomia \virg{Ettore e Majorana} dell' Universit\`{a} di Catania, Via S.Sofia 64, I-95123 Catania, Italy}
\address[add3]{\ Dipartimento di Fisica e Chimica - Emilio Segr\`{e}, Universit\`{a} degli Studi di Palermo, via Archirafi 36, I-90123 Palermo, Italy}

%

\begin{abstract}
We investigate the transfer of genuine multipartite entanglement across a spin-$\frac{1}{2}$ chain with nearest-neighbor $XX$-type interaction.
We focus on the perturbative regime, where a block of spins is weakly coupled at each edge of a quantum wire, embodying the role of a multiqubit sender and receiver, respectively. We find that high-quality multipartite entanglement transfer is achieved at the same time that three excitations are transferred to the opposite edge of the chain. Moreover, we find that both a finite concurrence and tripartite negativity is attained at much shorter time, making $GHZ$-distillation protocols feasible. Finally, we investigate the robustness of our protocol with respect to non-perturbative couplings and increasing lengths of the quantum wire.
\end{abstract}


\end{frontmatter}

\section{Introduction}\label{S.Intro}
Entanglement has become in the last few decades a central topic of many applications of quantum mechanics, ranging from quantum information~\cite{Nielsen:2011:QCQ:1972505} to quantum thermodynamics~\cite{10.1088/2053-2571/ab21c6}.
A great deal of work has been done to characterise its features and quantify the amount of entanglement shared among quantum systems, see, e.g., Ref.~\cite{Horodecki2009} and references therein, with the focus of defining measures related to entanglement resource theories~\cite{RevModPhys.91.025001}. 
However, apart from low-dimensional bipartite systems~\cite{PhysRevLett.77.1413}, there are no necessary and sufficient criteria to identify if a given quantum state is entangled, and the two qubits case is the only quantum system for which currently a complete characterization of its entanglement has been achieved for both pure and mixed states~\cite{doi:10.1063/1.1286032}. 
The multipartite entanglement in its simplest form, namely the tripartite entanglement shared among three qubits, is already so complex that no analytical expressions are know for its quantification. The main reason for this difficulty can be tracked back to the presence of two inequivalent SLOCC (stochastic local operations and classical communication) classes: the $GHZ$ and the $W$ class. Indeed, at variance with the two qubit scenario, where the Bell states represent the maximally entangled states and every other two qubit state can be generated from them, for three qubits, conversion between states belonging to the $GHZ$ and the $W$ class is impossible under SLOCC~\cite{PhysRevA.62.062314}.
Notwithstanding the conceptual and analytical difficulties related to three-qubit entanglement, the latter has found numerous applications both in fundamental physics, e.g., in experimental tests of non-locality without relying on Bell's inequality~\cite{Pan2000}, and in proposed quantum information processing protocols~\cite{universe5100209}, including cryptography~\cite{Hillery1999}, teleportation~\cite{PhysRevA.58.4394}, and quantum error correction~\cite{Reed2012}.

Besides the characterization and quantification of tripartite qubit entanglement, an important task is also its generation, distribution, and protection. A prototypical quantum channel is embodied by a quantum spin-$\frac{1}{2}$ chain, with the sender and receiver located at its edges~\cite{Bose2003a,doi:10.1142/S0217979213450355}. Whereas, the transfer of two-qubit entanglement has been extensively investigated via spin chain~\cite{Banchi2011a,Apollaro2015,Almeida2017, VIEIRA20182586,Vieira2019}, multipartite entanglement transfer between the edges of a spin chain has not yet been addressed. In this paper, we build on the perturbatively perfect excitations transfer scheme~\cite{chetcuti2019perturbativelyperfect} which has already been successfully adopted for one- and two-qubit quantum state transfer protocols~\cite{Wojcik2007,Lorenzo2016,qst2}. In Ref.~~\cite{chetcuti2019perturbativelyperfect}  it has been shown that three excitations can be transferred between the edges of a spin-$\frac{1}{2}$ $XX$-type chain provided that its length fulfills $N=4n+7$, with $n$ being a non-negative integer. The excitation transfer occurs in the weak-coupling regime, of the sender and receiver block to the wire, and it approaches unity in the limit of vanishing coupling, although at a price of a transfer time going to infinity. Such a regime has been dubbed perturbatively perfect (PP) excitation transfer. In this paper, we address the question as to whether the PP-excitation transfer protocol is also efficient for the transfer of tripartite entangled states, in particular for states belonging to the $GHZ$ class. In the limit of weak, but finite coupling the receiver state results to be in mixed state. Therefore, it is of the utmost importance to determine whether it is genuinely multipartite entangled and, eventually, be able to quantify its entanglement.    

The paper is organised as follows: in Sec.~\ref{S.Model} we introduce the model and its dynamics, in Sec.~\ref{S.3ent} we revise some of the available tripartite entanglement witnesses and monotones we will use for the analysis of the receiver state in Sec.~\ref{S.Ent_Trans}, where we report also the two-qubit concurrence. Finally, in Sec.~\ref{S.Conc} we draw conclusions and outlooks.

\section{The Model}\label{S.Model}
\begin{figure}[h!]
	\includegraphics[width=\linewidth]{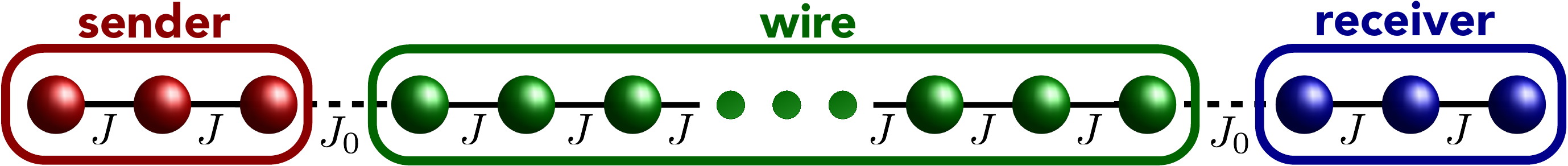}
	\caption{Setup of the excitation transfer protocol. A sender (red) and receiver (blue) block are weakly coupled by $J_0$ at both edges of a quantum wire (green). Each part is made up by a 1D lattice described by the Hamiltonian in Eq.~\ref{E.SpinXX} with $J_i=J=1$, and they are  coupled to each other by $J_0\ll 1$.}
		\label{F.model}
\end{figure}

In one of the most investigated quantum information transfer protocols, a sender is coupled to a receiver by means of a quantum wire~\cite{Bose2003a}. 
We rely on a similar set-up, where both the sender and the receiver are embodied by a block of three qubits, each weakly coupled to a quantum wire, see Fig.~\ref{F.model}.

We consider an $XX$ spin-$\frac{1}{2}$ Hamiltonian with nearest-neighbor interaction $J_i$ on a 1D lattice with open boundary conditions,
\begin{align}\label{E.SpinXX}
\hat{H} = \sum_{i=1}^{N-1}\frac{J_i}{2}\left(\hat{\sigma}^{x}_{i}\hat{\sigma}^{x}_{i+1} + \hat{\sigma}^{y}_{i}\hat{\sigma}^{y}_{i+1}\right)
\end{align}
where $\hat{\sigma}_i^{\alpha}$ ($\alpha=x,y$) represents the Pauli matrix of a spin-$\frac{1}{2}$ sitting on site $i$. We assume couplings $J_i$ all uniform but for $J_i=J_0$ between the sender (receiver) block and the wire. We also set the coupling within the sender (receiver) block and within the wire as our time and energy unit $J_i=J=1$.

By the Jordan-Wigner transformation, the Hamiltonian in Eq.~\ref{E.SpinXX} can be mapped~\cite{Lieb1961} into a spinless non-interacting fermion model,
\begin{align}
\label{E_Ham1}
\hat{H} = \sum\limits_{i=1}^{N-1}J_i\left(\hat{c}_{i+1}^{\dagger}\hat{c}_{i}+\hat{c}_{i}^{\dagger}c_{i+1}\right)~,
\end{align}
where the $\hat{c}_i^{\dagger}$ ($\hat{c}_i$) now represents a creation (distruction) operator of a spinless fermion on site $i$. Because of the quadratic nature of the Hamiltonian in Eq.~\ref{E_Ham1}, the diagonalisation is easily carried out and reads
\begin{align}\label{E_1part}
\hat{H}= \sum_{k=1}^N \omega_k \hat{c}_k^{\dagger}\hat{c}_k~,
\end{align}
where $\{\omega_k,\ket{\phi_k}\equiv \hat{c}_k^{\dagger}\ket{0}\}$ are the eigenvalues and the eigenvectors of the $N\text{x}N$ tridiagonal matrix with elements $\bra{i}\hat{H}\ket{j}=J_i\left(\delta_{i,i+1}+\delta_{i,i-1}\right)$, describing the single-particle dynamics in the direct space basis, $\ket{i}\equiv \hat{c}_i\ket{0}$. Here, and in the following, $\ket{i}\equiv\ket{00\dots 1_i\dots 00}$ represents a state with one excitation sitting on site $i$. 

As in the following we are interested in the transfer, from the sender to the receiver spins, of a $\ket{GHZ}$ state
\begin{align}
\label{E_GHZ}
\ket{GHZ}=\frac{1}{\sqrt{2}}\left(\ket{000}+\ket{111}\right)~,
\end{align}
we only need to explicate the dynamics in the 0- and 3-particle subspaces of Eq.~\ref{E_Ham1}, because the Hamiltonian in Eq.~\ref{E.SpinXX} conserves the total magnetisation in the $z$-direction. 

The dynamics in the 3-particle sector is fully determined by the transition amplitude matrix $\mathcal{F}_{ijk}^{nmr}(t)$ between sites $\{i,j,k\}$ and $\{n,m,r\}$, where $i<j<k$ and $n<m<r$, having single-particle transition amplitudes $f_s^r(t)$ as matrix elements~\cite{chetcuti2019perturbativelyperfect}

\begin{align}
\label{E.3ampl}
\mathcal{F}_{ijk}^{nmr}(t)= 
\bra{nmr}e^{-\imath \hat{H}t}\ket{ijk}=
\begin{pmatrix}
f_{i}^{n}(t)& f_{i}^{m}(t) &  f_{i}^{r}(t)\\
f_{j}^{n}(t) &f_{j}^{m}(t) & f_{j}^{r}(t)\\
f_{k}^{n}(t) &f_{k}^{m}(t) &  f_{k}^{r}(t)\\
\end{pmatrix}~.
\end{align}

The single-particle transition amplitude is given by
\begin{align}
\label{E.Sparticel_amp}
f_s^r(t)=\bra{r}e^{-it \hat{H}}\ket{s}=\sum_{k=1}^{N}e^{-i \omega_k t }\bra{r}\phi_k\rangle\!\langle \phi_k\ket{s}=\sum_{k=1}^{N}e^{-i \omega_k t }\phi_{rk}\phi^*_{ks}~,
\end{align}
evaluated via the eigenvalues and eigenstates of Eq.~\ref{E_1part}.
Finally, the square modulus of the determinant of the matrix in Eq.~\ref{E.3ampl} gives the transition probability of the excitations between the selected sites $\{i,j,k\}$ and $\{n,m,r\}$.
As for the 0-excitation sector, the fully polarised state $\ket{\mathbf{0}}\equiv \ket{00...0}$ is an eigenstate of the Hamiltonian whose evolution can be neglected by rescaling its eigenenergy to zero
\begin{align}
\label{E_evol}
\ket{\Psi(t)}=e^{-it \hat{H}}\ket{GHZ}_{123}\ket{\mathbf{0}}_{w,r}
=\frac{1}{\sqrt{2}}\left(\ket{\mathbf{0}}+e^{-it \hat{H}}\ket{111\mathbf{0}}\right)~,
\end{align}
where the initial state of our model consists of a $\ket{GHZ}$ state of the first three spins and all the spins of the wire and the receiver spins in the $\ket{0}$ state.

After a lengthy but straightforward calculation, the three qubits density matrix of the receiver block in the computational basis reads 
\begin{eqnarray}
\label{E.3qubit}
\rho_r(t)=
\begin{pmatrix}
	\rho_{00} & 0 & 0 & 0 & 0 & 0 & 0 & \rho_{07} \\ 
	0	& \rho_{11} & \rho_{12} & 0 & \rho_{14} & 0 & 0 & 0  \\ 
	0	& \rho_{12}^* & \rho_{22} & 0 & \rho_{24} & 0 & 0 & 0  \\ 
	0	& 0 & 0 & \rho_{33} & 0 & \rho_{35} & \rho_{36} & 0  \\ 
	0	& \rho_{14}^* & \rho_{24}^* & 0 & \rho_{44} & 0 & 0 & 0  \\ 
	0	& 0 & 0 & \rho_{35}^* & 0 & \rho_{55} & \rho_{56}&  0  \\ 
	0	& 0 & 0 & \rho_{36}^* & 0 & \rho_{56}^* & \rho_{66} & 0   \\ 
	\rho_{07}^*	& 0 & 0 & 0 & 0 & 0 & 0 & \rho_{77}
	\end{pmatrix}~.
\end{eqnarray}
Notice that the 1- and 2-excitations sector are block-diagonal as a consequence of the excitation-conserving property of the Hamiltonian, whereas the 0- and 3-particle sector are not because of the initial state of the sender. Each matrix element $\rho_{ij}$ can be expressed in terms of determinants of Eq.~\ref{E.3ampl} as follows
\begin{align}
\label{E_matrix_ele}
&\rho_{00}=\frac{1}{2}+\sum_{k<q<p=1}^{N-3}\left|\mathcal{F}_{123}^{kqp}\right|^2~,~\rho_{11}=\sum_{k=1}^{N-3}\left|\mathcal{F}_{123}^{N{-}2N{-}1k}\right|^2~,~\rho_{22}=\sum_{k=1}^{N-3}\left|\mathcal{F}_{123}^{N{-}2kN}\right|^2\nonumber\\
&\rho_{33}=\sum_{k<q=1}^{N-3}\left|\mathcal{F}_{123}^{N{-}2kq}\right|^2~,~
\rho_{44}=\sum_{k=1}^{N-3}\left|\mathcal{F}_{123}^{kN{-}1N}\right|^2~,~\rho_{55}=\sum_{k<q=1}^{N-3}\left|\mathcal{F}_{123}^{kN{-}1q}\right|^2\nonumber\\
&\rho_{66}=\sum_{k<q=1}^{N-3}\left|\mathcal{F}_{123}^{kqN}\right|^2~,~\rho_{77}=\left|\mathcal{F}_{123}^{N{-}2N{-}1N}\right|^2\nonumber\\
&\rho_{07}=\frac{1}{\sqrt{2}}\mathcal{F}_{123}^{N{-}2N{-}1N}~,~\rho_{12}=\sum_{k=1}^{N-3}\mathcal{F}_{123}^{N{-}2N{-}1k}\left(\mathcal{F}_{123}^{N{-}2kN}\right)^*~,~\rho_{14}=\sum_{k=1}^{N-3}\mathcal{F}_{123}^{N{-}2N{-}1k}\left(\mathcal{F}_{123}^{kN{-}1N}\right)^*~,~\nonumber\\
&\rho_{24}=\sum_{k=1}^{N-3}\mathcal{F}_{123}^{N{-}2kN}\left(\mathcal{F}_{123}^{kN{-}1N}\right)^*~,~\rho_{35}=\sum_{k<q=1}^{N-3}\mathcal{F}_{123}^{N{-}2kq}\left(\mathcal{F}_{123}^{kN{-}1q}\right)^*~,~\nonumber\\
&\rho_{36}=\sum_{k<q=1}^{N-3}\mathcal{F}_{123}^{N{-}2kq}\left(\mathcal{F}_{123}^{kqN}\right)^*~,~\rho_{56}=\sum_{k<q=1}^{N-3}\mathcal{F}_{123}^{kN{-}1q}\left(\mathcal{F}_{123}^{kqN}\right)^*~.
\end{align}

The reduced density matrix $\rho^{\left(ij\right)}$ of two qubits belonging to the receiver block is easily derived from Eq.~\ref{E.3qubit} and reads, e.g., for $\{ij\}=\{N{-}1,N\}$
\begin{align}
\label{E_2density_2}
\rho^{\left(N{-}1N\right)}=
\begin{pmatrix}
\rho_{00}+\rho_{11}& 0 & 0 & 0 \\ 
0 & \rho_{22}+\rho_{33} & \rho_{24}+\rho_{35} & 0 \\ 
& \rho_{24}^*+\rho_{35}^*  & \rho_{44}+\rho_{55} & 0 \\ 
0 & 0 & 0 & \rho_{66}+\rho_{77}
\end{pmatrix}~,
\end{align}
with a similar expression holding for the other pairs in the block. Notice that all the two-qubit density matrices are of $X$-type, and, consequently, the single-qubit density matrix will result diagonal.

Let us now recap some results from Ref.~\cite{chetcuti2019perturbativelyperfect} allowing us to express the single-particle transition amplitude of Eq.~\ref{E.Sparticel_amp} in terms of just a few eigenvectors of Eq.~\ref{E_1part} exploiting the perturbative coupling regime.
\begin{figure}[h!]
	\label{F.Figure_res}
	\includegraphics[width =\textwidth]{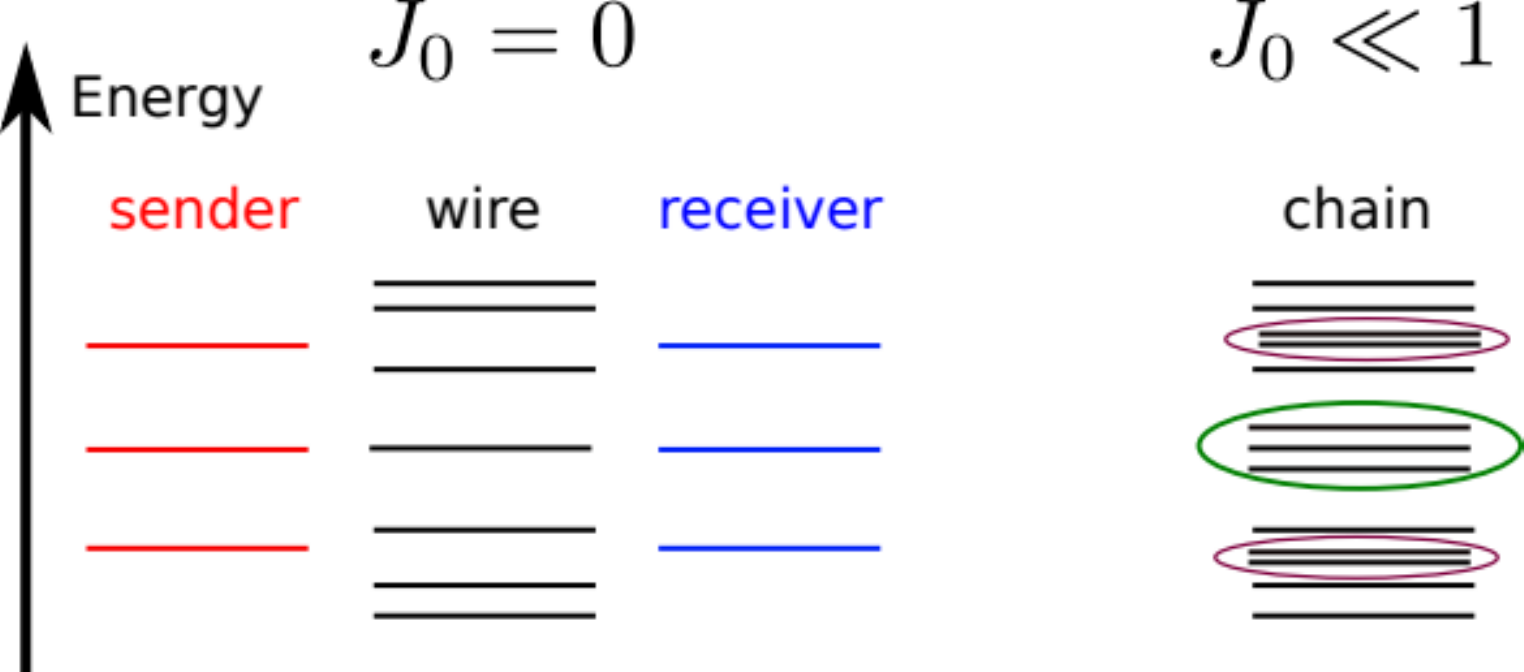}
	\caption{Perturbative analysis: the single-particle energy levels of Eq.~\ref{E_1part} when sender, receiver and wire are uncoupled (left panel) and in the weak-coupling regime (right panel). For lengths of the wire given by $n_w=4n+1$ there are one triple-degenerate level, resolved in energy at \nth{1}-order perturbation theory (green circled) and two symmetric double-degenerate levels, resolved at \nth{2}-order in $J_0$  (violet circle). Clearly, the energy separation of the former is of order $J_0^{-1}$, while the latter is of order $J_0^{-2}$. }
\end{figure}

In Fig.~\ref{F.Figure_res} we express graphically the effect of the perturbative coupling between the sender (receiver) block and the wire on the eigenenergies. It turns out that, for lengths of the wire obeying $n_w=4n+1$, with $n$ a non-negative integer, there are three resonances,  one is at \nth{1}-order in perturbation theory, and two are at \nth{2}-order, symmetrically displaced around the former. As a consequence, only seven eigenstates give a perturbatively non-negligible contribution to Eq.~\ref{E.Sparticel_amp}, which can be reduced to just three taking into account the mirror-symmetry of the model, reflected by the symmetrical displacement of the \nth{2}-order perturbed eigenenergies.
Utilising elementary trigonometric identities, each single-particle transition amplitude entering Eqs.~\ref{E_matrix_ele} via Eq.~\ref{E.3ampl} is a function of only three frequencies $\omega_{76}^{\pm}=\frac{\omega_7\pm\omega_6}{2}$ and $\omega_5$. The former are the eigenenergies corrected at \nth{2}-order in perturbation theory, the latter at \nth{1}-order, corresponding to the circles eigenergies in Fig.~\ref{F.Figure_res} ordered from below.
For instance, the transition amplitude of an excitation between site 1 and $N{-}2$ reads
\begin{align}
\label{E.Sparticel_amp_2}
f_1^{N{-}2}(t)\simeq\sum_{k=1}^{3}e^{-i \omega_k t }\phi_{rk}\phi^*_{ks}=\frac{1-2 \sin\omega_{67}^+\sin\omega_{67}^--\cos\omega_{5} }{4}~,
\end{align}
and between site 2 and $N{-}1$
\begin{align}
\label{E.Sparticel_amp_3}
f_2^{N{-}1}(t)\simeq\sum_{k=1}^{3}e^{-i \omega_k t }\phi_{rk}\phi^*_{ks}=-\sin\omega_{67}^+\sin\omega_{67}^-~.
\end{align}

Because of the different perturbation order corrections, the relation $\omega_{76}^-\ll\omega_5\ll\omega_{76}^+$ holds. This, as we will see, gives rise to two different time scales, $T\simeq\frac{\pi}{2 \omega_{76}^-}$ and $\widetilde{T}\simeq\frac{\pi}{2 \omega_5}$ dominating the oscillatory behaviour of the entanglement under scrutiny in the following sections.
Let us also stress that the \nth{1}-order doublet and the two \nth{2}-order triplet perturbed eigenenergy each can support only one excitation.

\section{Three qubit entanglement}\label{S.3ent}
Having derived in the previous section~\ref{S.Model} the tools to obtain the receiver three qubit density matrix, in this section we will overview a few results about multipartite entanglement we will use to tackle the multipartite entanglement transfer problem.

Whereas two qubit entanglement criteria for an arbitrary density matrix have been derived~\cite{PhysRevLett.77.1413} and entanglement monotones have a closed expression~\cite{PhysRevLett.80.2245}, for the entanglement shared among three qubits the scenario is much more complex, and, for arbitrary mixed states no closed expression of an entanglement measure is known. 

One of the difficulties in characterising the entanglement shared among three qubits is the existence  of six
different  SLOCC (stochastic local operations and classical communication) classes for pure states: the $GHZ$- and $W$-class for genuinely entangled states, three classes are composed by a two-qubit Bell state and single qubit state embodying the bi-separable states with respect to each possible partition, and, finally, a product state of three qubits representing the fully separable state~\cite{PhysRevA.62.062314}. 

\begin{figure}[h!]
	\label{F_3qubit_class}
	\includegraphics[width =\textwidth]{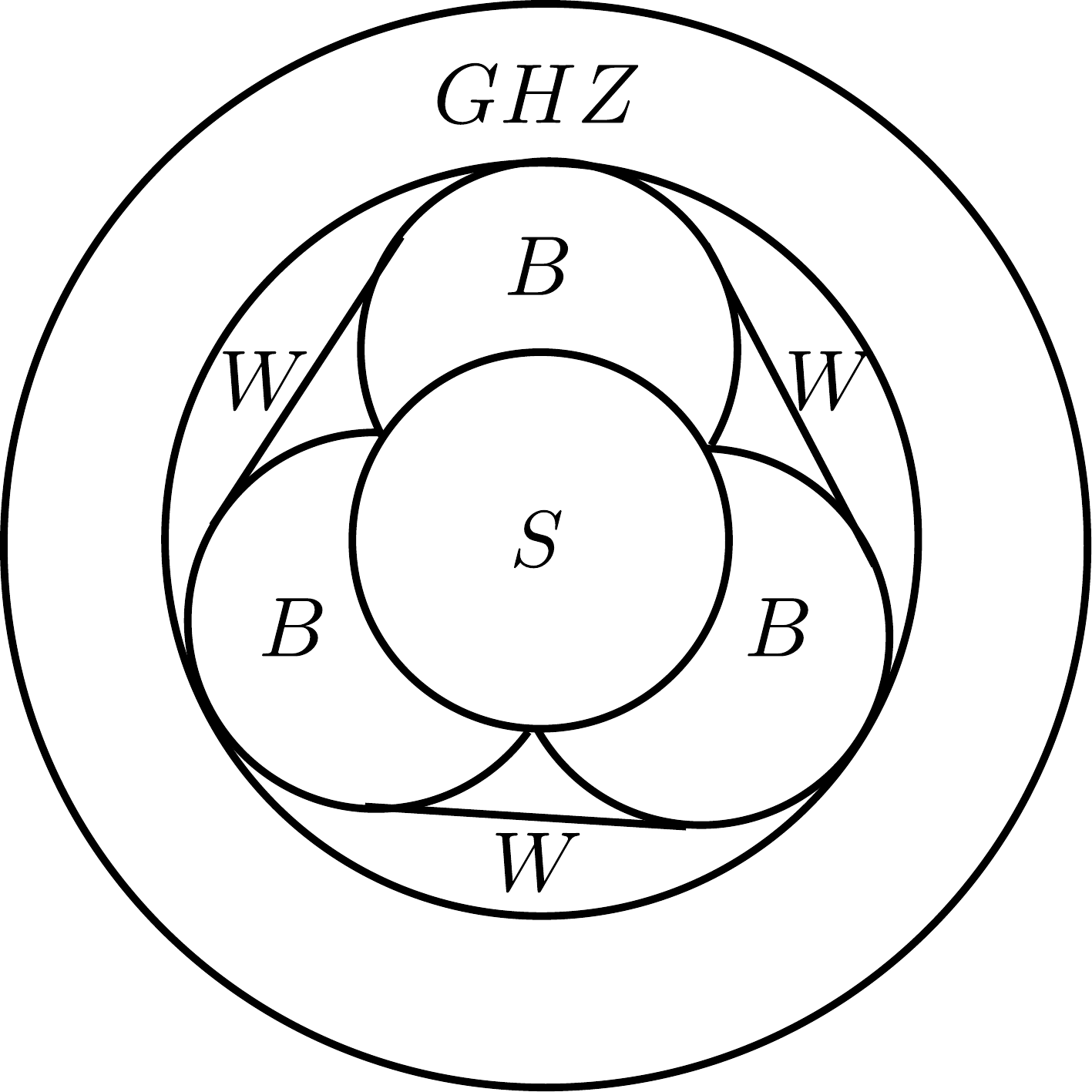}
	\caption{Schematic diagram of the classification of three qubit states: $S$ fully separable, $B$ bi-separable, $W$ and $GHZ$ non-separable.}
\end{figure}

This classification has been extended to mixed states~\cite{Acin2001a} giving rise to a hierarchy of entanglement where local POVMs can transform states only from a higher to a lower class, whereas each class is invariant under SLOCC, see Fig.\ref{F_3qubit_class} for the schematic structure. However, while pure states that are biseparable with respect to each partition are also fully separable, the same does not hold for mixed states because of the existence of PPT entangled states.

For three qubits a pure state is called fully separable if it can be written in the form 
\begin{align}
\label{E.3purefs}
\ket{\Psi_{fs}}=\ket{\psi_1}\ket{\psi_2}\ket{\psi_3}~
\end{align}  
and a mixed state belongs to the fully separable class $S$ if it can be written as a convex combination of fully separable pure states
\begin{align}
\label{E.3mixedfs}
\ket{\rho_{fs}}=\sum_i p_i  \ket{\Psi_{fs}}\!\!\bra{\Psi_{fs}}~.
\end{align}
A pure bi-separable state, belonging to the class $B$, is defined as being separable under one, or more, bi-partitions, $\{1|23,12|3,13|2\}$, as, e.g., in
\begin{align}
\label{E.3purebs}
\ket{\Psi_{bs}}=\ket{\psi_{12}}\ket{\psi_3}~,
\end{align} 
with qubits 1 and 2 possibly entangled. Consequently, a bi-separable mixed state reads
\begin{align}
\label{E.3mixedbs}
\ket{\rho_{bs}}=\sum_i p_i  \ket{\Psi_{bs}}\!\!\bra{\Psi_{bs}}~.
\end{align}
If a mixed state can not be written as in Eqs.~\ref{E.3mixedfs} or ~\ref{E.3mixedbs}, it contains genuine multipartite entanglement, which can be of the $W$ or of the $GHZ$-type. It holds that $S\subset B\subset W\subset GHZ$~\cite{Acin2001a}. 

In order to determine to which SLOCC class a three qubit pure state belongs, one can rely on the three-tangle $\tau$~\cite{PhysRevA.61.052306} and the concurrence $C$. The $GHZ$ class contains states with $\tau>0$, whereas states in the $W$ class have   $\tau=0$ but finite $C^{\left(12\right)}$, $C^{\left(13\right)}$, and $C^{\left(23\right)}$; bi-separable states have only one of the above concurrences different from zero, and, finally, for fully separable states both $\tau$ and all $C^{\left(ij\right)}$ vanish.
This classification extends to mixed states by considering the classes in its pure state decomposition and using the convex roof extension of a corresponding pure state entanglement measure: for $GHZ$-type entanglement $\tau$ is finite, for $W$-type entanglement $\tau=0$ but the concurrence of genuine multipartite entanglement~\cite{PhysRevA.83.062325} is finite; whereas, for bi-separability, the square root of the global entanglement~\cite{doi:10.1063/1.1497700} is finite, and both the three-tangle and concurrence of genuine multipartite entanglement are zero. Finally, for states in the fully separable class, all entanglement measures vanish. 

Generally, convex roof extensions of pure state entanglement measures are difficult to calculate as they involve an optimisation over an infinite number of convex decompositions into pure states of a mixed state. Although efficient numerical algorithms have been developed for several multipartite entanglement measures, see, e.g.,Ref.~\cite{PhysRevLett.114.160501}, for full rank density matrices, as is the one given by Eq.~\ref{E.3qubit}, there is no efficient algorithm available to date.

An alternative to entanglement measures is given by entanglement witnesses (EW)~\cite{TERHAL2000319}. An EW is an hermitian operator $\mathcal{W}$ such that $Tr\left[\mathcal{W}\rho\right]\geq 0$ on all states $\rho$ not belonging to the entanglement class the EW aims at detecting. As such, $\mathcal{W}$ is a witness in the sense it constitutes a sufficient, but not necessary criterion for detecting entanglement. 
For the GHZ-class several witnesses have been devised and their decomposition into local projective measurements have allowed to detect experimentally genuine multipartite entanglement~\cite{PhysRevLett.92.087902}.

In our analysis of the transfer of multipartite entanglement, we will use the entanglement witnesses of Ref.~\cite{Acin2001a}
\begin{eqnarray}
\label{E.EW_GHZ}
\mathcal{W}=\frac{1}{2}\mathbb{1}-\opkb{GHZ}{GHZ}~.
\end{eqnarray}
One has $\Trace{\mathcal{W}\rho}>0$ on every biseparable state of Eq.~\ref{E.3mixedbs}, for $-\frac{1}{4}<\Trace{\mathcal{W}\rho}<0$ the state $\rho$ can belong either to the $W$ or the $GHZ$-class, 
while only states belonging to the $GHZ$ class have $-\frac{1}{2}<\Trace{\mathcal{W}\rho}<-\frac{1}{4}$. 
For states belonging to the $W \setminus B$ class, the following witness can be used $\mathcal{W}_W=\frac{2}{3}\mathbb{1}-\opkb{W}{W}$.

In Ref.~\cite{Jungnitsch2011} a semidefinite programming (SDP) approach has been put forward in order to detect multipartite entanglement, although without distinguishing between the $GHZ$- and the $W$-type entanglement. Using convex optimisation technique, one is able to solve, for an arbitrary multipartite state $\rho$, the minimization problem
\begin{align}
\label{E_SDP}
\min \Trace{\mathrm{W} \rho}~,
\end{align}
where $\mathrm{W}$ is a fully decomposable witness with respect to every bipartition of the multipartite system. Interestingly, (the negative of) Eq.~\ref{E_SDP} is also a multipartite entanglement monotone and can hence be used to quantify genuine multipartite entanglement~\cite{PhysRevA.99.012319}.

Apart from entanglement witnesses, the quantification of entanglement in a three qubit mixed state via the tangle $\tau$ is possible only in a few specific low-rank cases~\cite{Eltschka_2014}. However, bipartite entanglement measures can be used on multipartite states by considering every possible partitions~\cite{YU2004377}, and we will use in the following the tripartite negativity $N_{ABC}$ proposed in Ref.~\cite{Sabin2008}:
\begin{align}
\label{E_tripart_neg}
N^{(3)}=\sqrt[3]{N_{A|BC}N_{AB|C}N_{AC|B}}~
\end{align}
where $N_{X|YZ}$ is the negativity~\cite{Vidal2002a} 
\begin{align}
\label{E_neg3}
N_{X|YZ}=\frac{\sum_i \left|\lambda\right|-1}{2}~,
\end{align}
 with $\lambda$ being the eigenvalues of the partial transpose of $\rho_{XYZ}$ with respect to the subsystem $X$. 
 However, due to the Peres-Horodecki criterion~\cite{HORODECKI19961, PhysRevLett.77.1413}, for dimensions higher than $2\text{x}2$ and $2\text{x}3$, $N_{X|YZ}>0$ constitutes a sufficient, but not necessary condition for bipartite entanglement between the partitions $X$ and $YZ$. Notice that $N^{(3)}\left(\rho\right)>0$ is a sufficient condition for distillability of a $GHZ$ state from $\rho$~\cite{Dur1999}.
 
Finally, let us also report for completeness, the concurrence between two qubits $i$ and $j$, $C^{(ij)}$,~\cite{PhysRevLett.80.2245}. Because all the two-qubit density matrices $\rho^{(ij)}$ in Eq.~\ref{E_2density_2} are of $X$-type, with a single non-zero off-diagonal element, the concurrence reduces to~\cite{Amico2004b}
\begin{align}
\label{E.Conc_mio}
C^{(ij)}=2 \max\left[0,\left|\rho^{(ij)}_{12}\right|-\sqrt{\rho^{(ij)}_{00}\rho^{(ij)}_{33}}\right]~.
\end{align} 
\subsection{Entanglement transfer}\label{S.Ent_Trans}
\begin{figure}[h!]
	\label{F.3ent}
	\includegraphics[width =\textwidth]{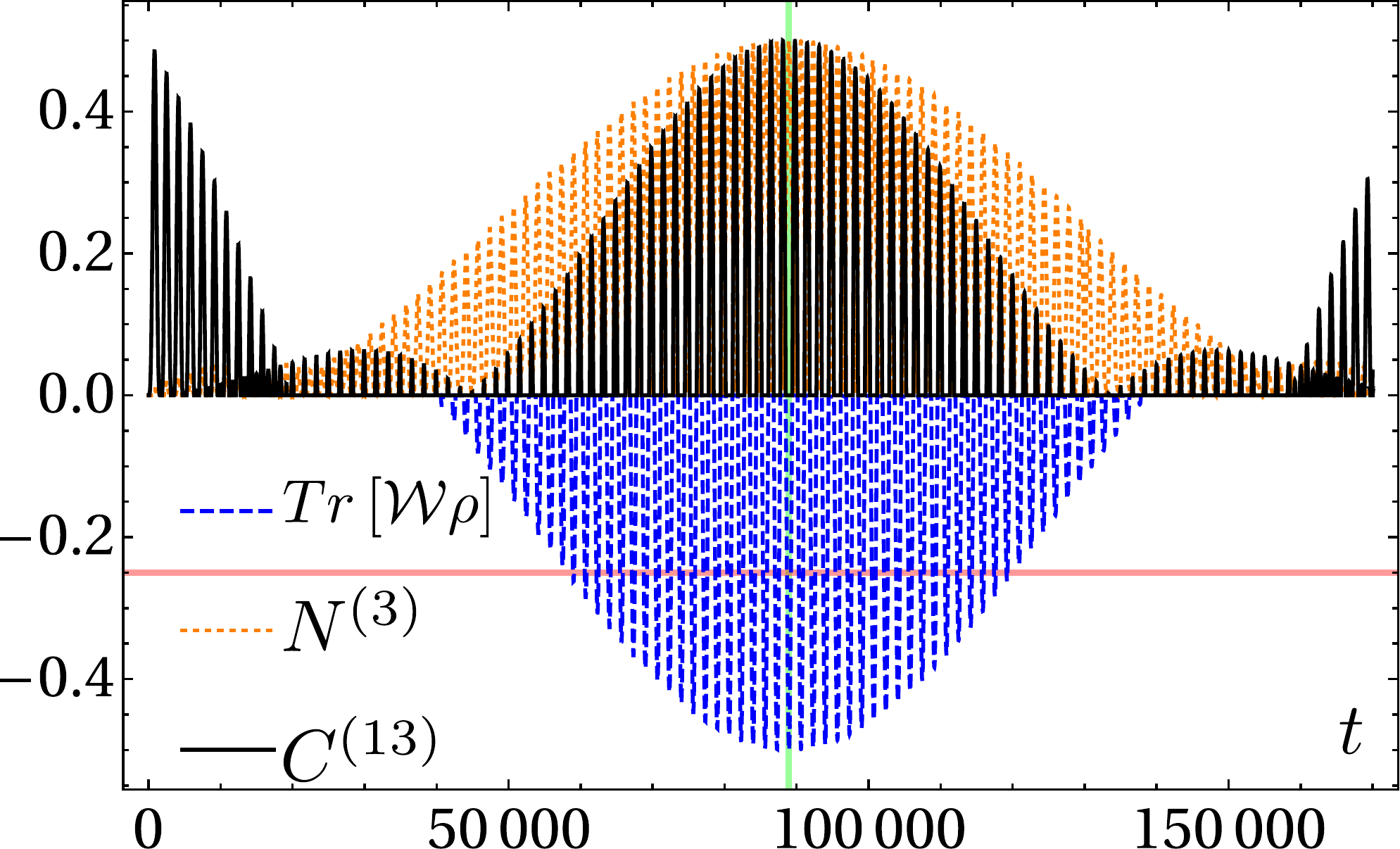}
	\includegraphics[width =\textwidth]{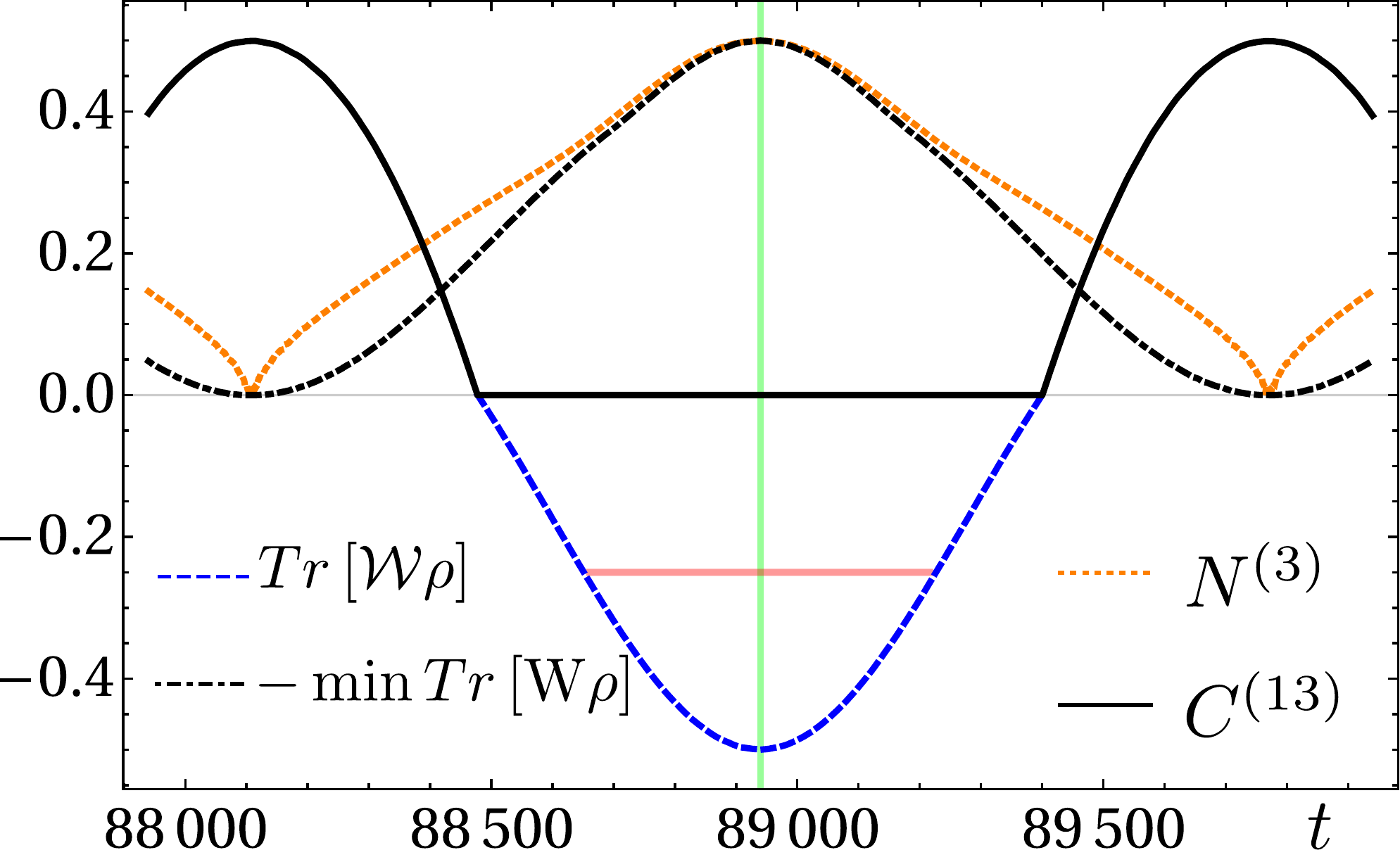}
	\caption{ Witness
		$\mathcal{W}$, Eq~\ref{E.EW_GHZ}, (blue dotted line) and $\mathrm{W}$, Eq.~\ref{E_SDP}, (black dotted line); tripartite negativity $N^{(3)}$, Eq.~\ref{E_tripart_neg}, (orange dotted line), and  concurrence between qubit 1 and 3, $C^{\left(13\right)}$, Eq~\ref{E.Conc_mio}, (blue line) on a time scale of $T=\frac{2\pi}{\omega_{76}^-}$ (upper panel) and a few  $\widetilde{T}=\frac{2\pi}{\omega_{5}}$ (lower panel) around the maximum of the fidelity given by $\tau=\frac{T}{2}$ (green vertical line). The horizontal red line is set at $-\frac{1}{4}$ to detect $GHZ$-class entanglement via the witness $\mathcal{W}$. Notice that, around $t=\tau$, $C^{\left(13\right)}$ and $N^{(3)}$ are oscillating in phase opposition.}
\end{figure}

Let us now finally illustrate the main results of this work: the transfer of multipartite entanglement via perturbative couplings between a sender and a receiver block connected by a quantum wire. As we are interested only in the receiver block, we renumber, for the sake of readability, the spins therein contained $n=1,2,3$, starting from the edge. 
In Fig.~\ref{F.3ent} we report the results for two entanglement witnesses, respectively given by Eq.~\ref{E.EW_GHZ}
and Eq.~\ref{E_SDP}, the tripartite negativity $N^{(3)}$, Eq.~\ref{E_neg3}, and the concurrence $C^{\left(13\right)}$, Eq~\ref{E.Conc_mio}, between qubit 1 and 3 for a chain of length $N=19$ and $J_0=0.01$ both on a time scale of $T$ and $\widetilde{T}$. Being the concurrence between neighboring qubits $C^{\left(12\right)}=C^{\left(23\right)}=0$, and the witness $\mathcal{W}_W$ detecting $W$-class states positive at all times when evaluated on the receiver density matrix, $\Trace{\mathcal{W}\rho_r}>0$, we argue that no $W$-entanglement is present at any time in the receiver spins.

Whereas the witness based on the fidelity with a $GHZ$ state, Eq.~\ref{E.EW_GHZ}, detects genuine multipartite entanglement at finite time-intervals, the witness in Eq.~\ref{E_SDP} detects genuine multipartite entanglement at any time but for discrete time points. The latter coincide with the times when the tripartite negativity $N^{\left(3\right)}$ vanishes. There are regions where the witness $-\frac{1}{4}<\Trace{\mathcal{W\rho}}<0$, hence the tripartite entanglement could be either $GHZ$ or $W$-type. However, the fact that $C^{\left(12\right)}=C^{\left(23\right)}=0$ is an indication that $\rho$ belongs to the $GHZ$-class. The same holds for the regions where the multipartite entanglement monotone derived Eq.~\ref{E_SDP} gives a non-zero value.

We also observe that both $N^{(3)}$ and $-\Trace{\mathrm{W}\rho}$ oscillate with period $\widetilde{T}$, becoming vanishingly small when the concurrence between qubit 1 and 3 reaches its maximum value, $C^{\left(13\right)}=\frac{1}{2}$.
At these points $t^*$ in time, the density matrix of the receiver block reads
\begin{align}
\label{E_den3_sep}
\rho(t^*)&\simeq \frac{1}{2}\opkb{000}{000}+\frac{1}{2}\left(\frac{\ket{110}-\ket{011}}{\sqrt{2}}\right)\!\left(\frac{\bra{110}-\bra{011}}{\sqrt{2}}\right)\\
&=\frac{1}{2}\opkb{0_A0_C}{0_A0_C}\otimes\opkb{0_B}{0_B}\nonumber\\
&+\frac{1}{2}\left(\frac{\ket{1_A0_C}-\ket{0_A1_C}}{\sqrt{2}}\right)\!\left(\frac{\bra{1_A0_C}-\bra{0_A1_C}}{\sqrt{2}}\right)\otimes\opkb{1_B}{1_B}\nonumber~,
\end{align}
which is a biseparable state under the partition $AC|B$. Therefore, we can conclude that these are the only (isolated) points in time where the state does not have any genuinely multipartite entanglement. Clearly, the reason for these oscillations is that one of the excitations is travelling with frequency $\omega_5$ back and forth between the sender and the receiver block through the quantum wire exploiting the \nth{1}-order triplet. 

\begin{figure}[h!]
	\label{F3ent2}
	\includegraphics[width =0.5\textwidth]{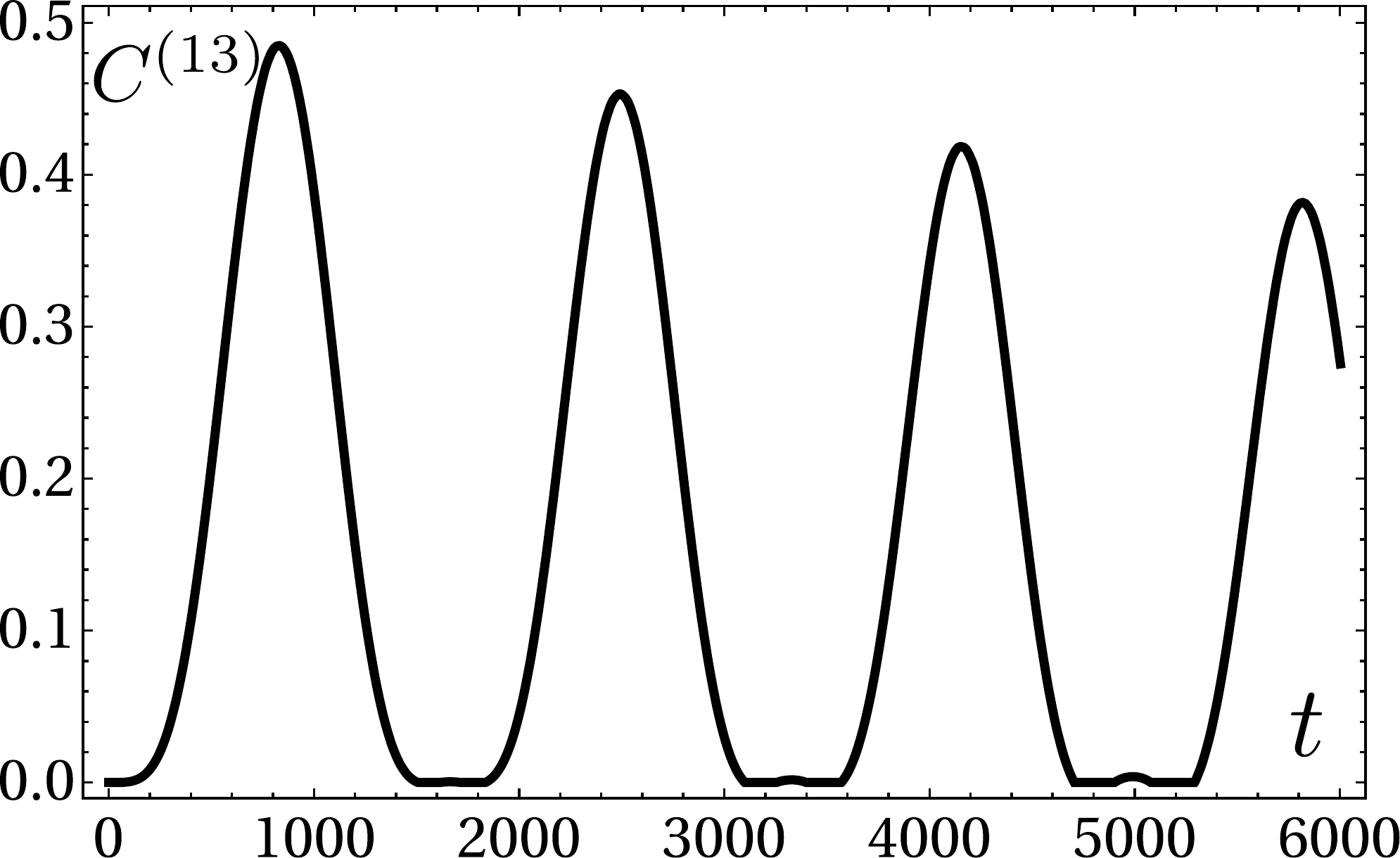}
	\includegraphics[width =0.5\textwidth]{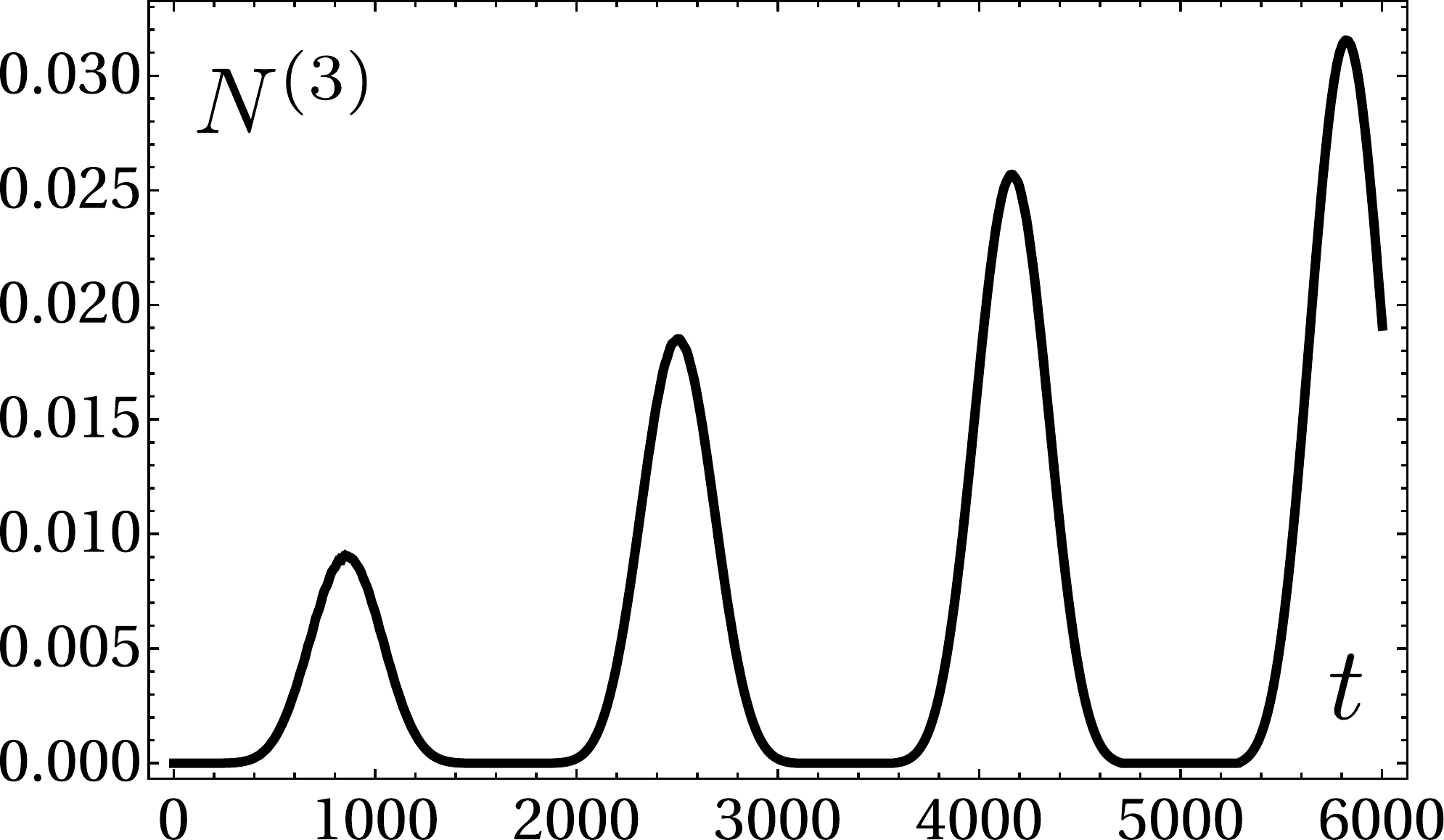}
	\caption{Short-time behaviour of the concurrence $C^{\left(13\right)}$, Eq~\ref{E.Conc_mio}, (left panel) and of the tripartite negativity $N^{(3)}$, Eq.~\ref{E_tripart_neg} (right panel). Notice that, at variance with the time scale of $T$, on a times scale of the order of $\widetilde{T}$, the two entanglement quantifiers are oscillating in phase.}
\end{figure}

Analysing the short-time behaviour, we notice that qubits 1 and 3 get entangled with $C^{\left(13\right)}=\frac{1}{2}$ already on a time-scale of $\widetilde{T}$, whereas the tripartite negativity $N^{(3)}$, as well as the entanglement monotone $-\Trace{\mathrm{W}\rho}$, is very small, Fig.~\ref{F3ent2}. 
The reason still being the presence of the \nth{1}-order triplet, entering the transition amplitudes $f_i^j$ with $i=1,3$ and $j=N{-}2,N$. Whereas, in order to have finite genuinely tripartite entanglement one needs a finite probability to find three excitations on the receiver block, thus involving the two \nth{2}-order doublets, which is the only term entering the transition amplitude in Eq.~\ref{E.Sparticel_amp_3}.

\begin{figure}[h!]
	\label{F3ent1}
	\includegraphics[width =0.5\textwidth]{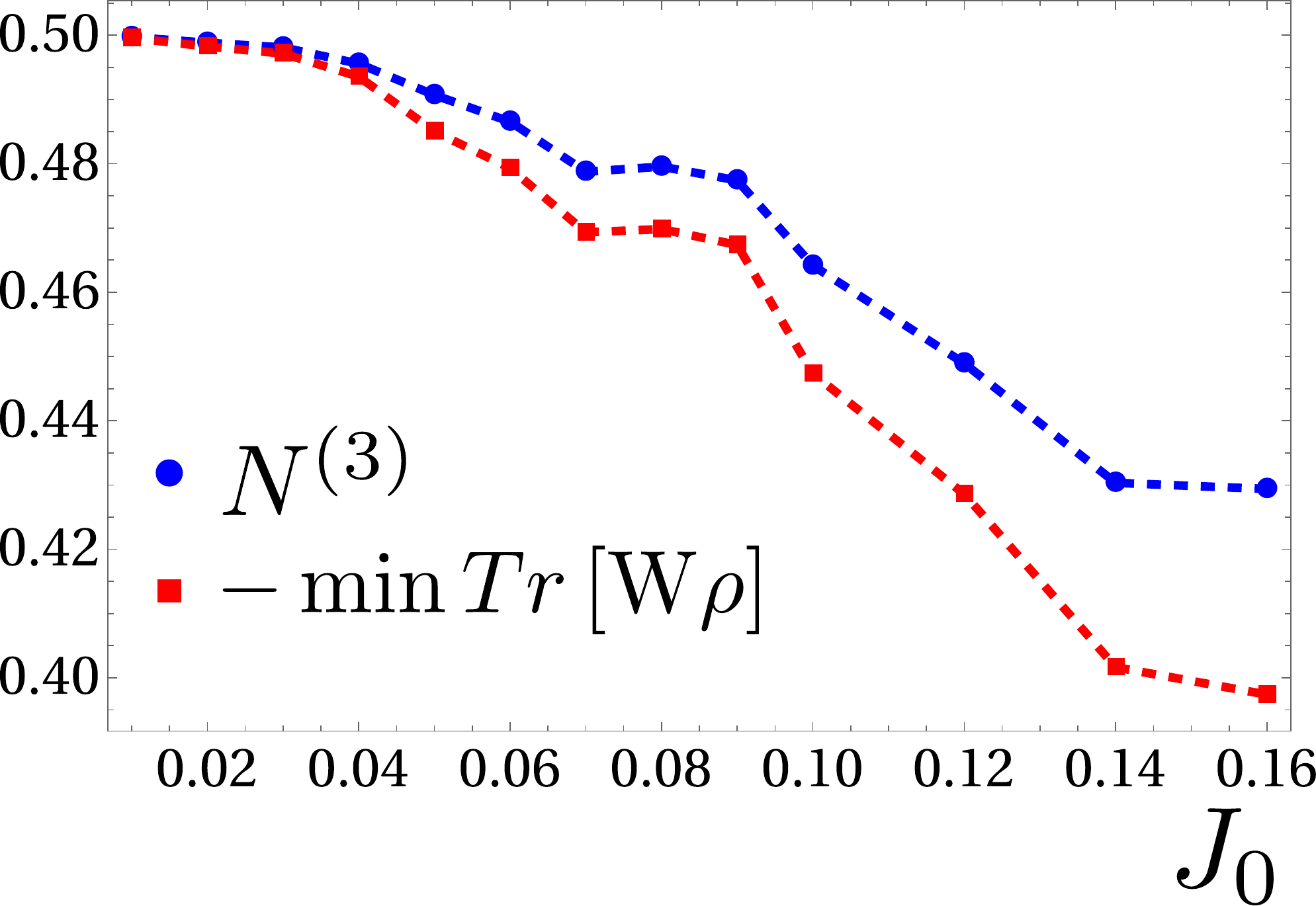}
	\includegraphics[width =0.5\textwidth]{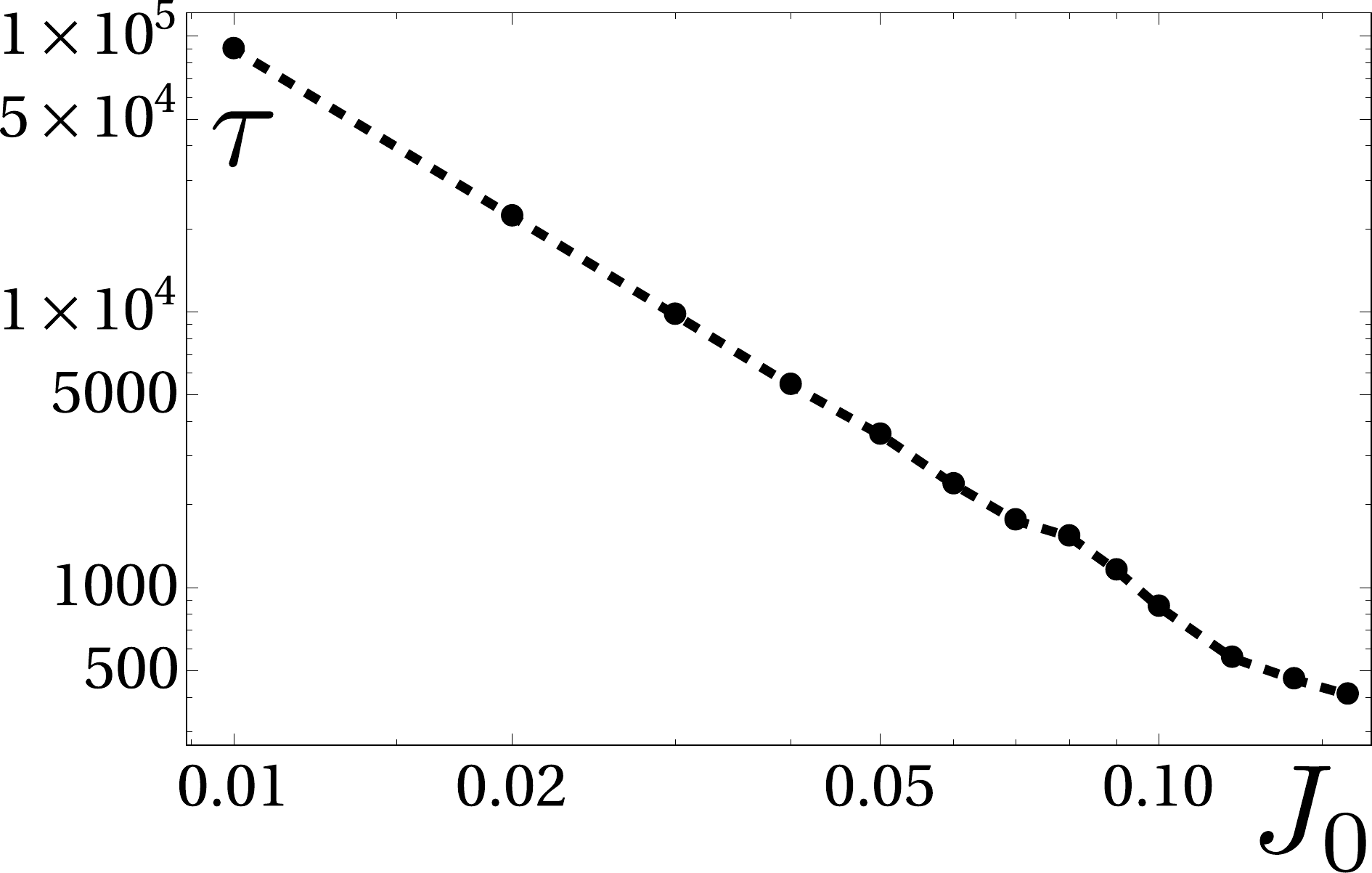}
	\caption{
		(left panel) Maximum tripartite negativity $N^{\left(3\right)}$, Eq.~\ref{E_tripart_neg}, blue dots, and witness $\mathrm{W}$, Eq.~\ref{E_SDP}, red square, as a function of the coupling $J_0$ for a chain of length $N=23$.  (right panel) Time $\tau$ at which the maximum are achieved vs $J_0$ for the same parameters as in the left panel. Notice how the two entanglement monotones change outside the weak-coupling regime. Similarly, around the same values, the power law $\tau\sim J_0^{-2}$, obtained from \nth{2}-order perturbation theory in Ref.~\cite{chetcuti2019perturbativelyperfect}, starts to fail. Lines are for guiding the eyes.}
\end{figure}

In Fig.~\ref{F3ent1} we test our protocol for increasing values of $J_0$ and report a good transfer of genuine multipartite entanglement to the sender block in the weak-couling regime, say up to $J_0\simeq 0.1$, after which a quick decay of the quality of the transfer is observed. Similarly, the transfer time $\tau$ at which the maximum is obtained follows $\tau\propto J_0^{-2}$ in the perturbative regime, before breaking down after $J_0\simeq 0.1$.

Let us also add that $C_{\#}^{(13)}=C^{\left(13\right)}=\frac{1}{2}$, where $C_{\#}^{(13)}$ is the concurrence of assistance~\cite{DBLP:journals/qic/LaustsenVE03}, evaluated by
	\begin{align}
	\label{E.Conc_ass}
	C_{\#}^{(ij)}=\sum_{n=1}^{4}\sqrt{\lambda_n}~,
	\end{align}
where $\lambda_n$ are the eigenvalues of the matrix $R=\rho \left(\hat{\sigma}^y\otimes\hat{\sigma}^y\right)\rho^*\left(\hat{\sigma}^y\otimes\hat{\sigma}^y\right)$. This quantity is the maximum entanglement achievable between two qubits
by means of LOCC operations on the complementary qubits, that is the sender and the wire qubits. 

\section{Conclusions}\label{S.Conc}
In this paper we have shown how, by exploiting the weak-coupling dynamical regime, one is able to transfer maximally entangled three-qubit states between the edges of a spin-$\frac{1}{2}$ chain with nearest-neighbor $XX$-interactions. We have used a witness based on the fidelity with a $GHZ$ state and one based on semidefinite programming. The negative of the latter, as well as the tripartite negativity, constitute also valid entanglement monotones and their dynamics shows that genuine multipartite entanglement of the $GHZ$-class is efficiently transferred with our protocol.
Interestingly, although the multipartite entanglement transfer peaks at a time scale determined by the inverse of the \nth{2}-order energy gap in perturbation theory, a finite concurrence between a pair of spins in the receiver block and a non-zero tripartite negativity, the latter a sufficient condition for $GHZ$-distillability, is retrieved on the sender block already on times scales determined by the much faster inverse of the \nth{1}-order energy gap in perturbation theory. 
 Moreover, while in the limit of vanishing couplings of the sender and the receiver block to the quantum wire, $J_0\rightarrow 0$, the transfer of the multipartite entanglement approaches one, but with the transfer time approaching infinity, 
 we obtain that also for couplings $J_0\sim 0.1$, a significative amount of multipartite entanglement is retrieved on the receiver spins in much shorter time.
Finally, although we were not able to evaluate the tangle, because of the full rank of the receivers density matrix, both the witnesses and the entanglement monotones considered indicate that for a large time interval around the transfer time, the receivers state remains genuinely multipartite entangled, but for isolated points in time.

Taking into consideration that genuine multipartite entanglement, despite the analytical, and even numerical, difficulty of its characterisation and quantification, is a precious resource in many applications, ranging from cryptography to quantum error correction, we believe that a thorough investigation of its dynamical properties may result useful and more studies in this direction are needed.

\section*{Ackowledgements}
The authors thank Prof.~Andr\'{e} Xuereb and Dr.~Zsolt Bern\'{a}d for useful discussions. C. S. acknowledges funding by the European Union's Horizon 2020 research and innovation programme under Grant Agreement No.\ 732894 (FET Proactive HOT).
\section*{References}

\bibliography{mybibfile}

\begin{thebibliography}{10}
\expandafter\ifx\csname url\endcsname\relax
  \def\url#1{\texttt{#1}}\fi
\expandafter\ifx\csname urlprefix\endcsname\relax\def\urlprefix{URL }\fi
\expandafter\ifx\csname href\endcsname\relax
  \def\href#1#2{#2} \def\path#1{#1}\fi

\bibitem{Nielsen:2011:QCQ:1972505}
M.~A. Nielsen, I.~L. Chuang, Quantum Computation and Quantum Information: 10th
  Anniversary Edition, 10th Edition, Cambridge University Press, New York, NY,
  USA, 2011 (2011).

\bibitem{10.1088/2053-2571/ab21c6}
S.~Deffner, S.~Campbell,
  \href{http://dx.doi.org/10.1088/2053-2571/ab21c6}{Quantum Thermodynamics},
  2053-2571, Morgan \& Claypool Publishers, 2019 (2019).
\newblock \href {https://doi.org/10.1088/2053-2571/ab21c6}
  {\path{doi:10.1088/2053-2571/ab21c6}}.
\newline\urlprefix\url{http://dx.doi.org/10.1088/2053-2571/ab21c6}

\bibitem{Horodecki2009}
R.~Horodecki, P.~Horodecki, M.~Horodecki, K.~Horodecki,
  \href{https://link.aps.org/doi/10.1103/RevModPhys.81.865}{{Quantum
  entanglement}}, Reviews of Modern Physics 81~(2)  865--942 (jun).
\newblock \href {https://doi.org/10.1103/RevModPhys.81.865}
  {\path{doi:10.1103/RevModPhys.81.865}}.
\newline\urlprefix\url{https://link.aps.org/doi/10.1103/RevModPhys.81.865}

\bibitem{RevModPhys.91.025001}
E.~Chitambar, G.~Gour,
  \href{https://link.aps.org/doi/10.1103/RevModPhys.91.025001}{Quantum resource
  theories}, Rev. Mod. Phys. 91 (2019) 025001 (Apr 2019).
\newblock \href {https://doi.org/10.1103/RevModPhys.91.025001}
  {\path{doi:10.1103/RevModPhys.91.025001}}.
\newline\urlprefix\url{https://link.aps.org/doi/10.1103/RevModPhys.91.025001}

\bibitem{PhysRevLett.77.1413}
A.~Peres,
  \href{https://link.aps.org/doi/10.1103/PhysRevLett.77.1413}{Separability
  criterion for density matrices}, Phys. Rev. Lett. 77 (1996) 1413--1415 (Aug
  1996).
\newblock \href {https://doi.org/10.1103/PhysRevLett.77.1413}
  {\path{doi:10.1103/PhysRevLett.77.1413}}.
\newline\urlprefix\url{https://link.aps.org/doi/10.1103/PhysRevLett.77.1413}

\bibitem{doi:10.1063/1.1286032}
K.~G.~H. Vollbrecht, R.~F. Werner,
  \href{https://aip.scitation.org/doi/abs/10.1063/1.1286032}{Why two qubits are
  special}, Journal of Mathematical Physics 41~(10) (2000) 6772--6782 (2000).
\newblock \href
  {http://arxiv.org/abs/https://aip.scitation.org/doi/pdf/10.1063/1.1286032}
  {\path{arXiv:https://aip.scitation.org/doi/pdf/10.1063/1.1286032}}, \href
  {https://doi.org/10.1063/1.1286032} {\path{doi:10.1063/1.1286032}}.
\newline\urlprefix\url{https://aip.scitation.org/doi/abs/10.1063/1.1286032}

\bibitem{PhysRevA.62.062314}
W.~D\"ur, G.~Vidal, J.~I. Cirac,
  \href{https://link.aps.org/doi/10.1103/PhysRevA.62.062314}{Three qubits can
  be entangled in two inequivalent ways}, Phys. Rev. A 62 (2000) 062314 (Nov
  2000).
\newblock \href {https://doi.org/10.1103/PhysRevA.62.062314}
  {\path{doi:10.1103/PhysRevA.62.062314}}.
\newline\urlprefix\url{https://link.aps.org/doi/10.1103/PhysRevA.62.062314}

\bibitem{Pan2000}
J.-W. Pan, D.~Bouwmeester, M.~Daniell, H.~Weinfurter, A.~Zeilinger,
  \href{https://doi.org/10.1038/35000514}{Experimental test of quantum
  nonlocality in three-photon greenberger-horne-zeilinger entanglement}, Nature
  403~(6769) (2000) 515--519 (2000).
\newblock \href {https://doi.org/10.1038/35000514}
  {\path{doi:10.1038/35000514}}.
\newline\urlprefix\url{https://doi.org/10.1038/35000514}

\bibitem{universe5100209}
M.~M.~Cunha, A.~Fonseca, E.~O.~Silva,
  \href{https://www.mdpi.com/2218-1997/5/10/209}{Tripartite entanglement:
  Foundations and applications}, Universe 5~(10) (2019).
\newblock \href {https://doi.org/10.3390/universe5100209}
  {\path{doi:10.3390/universe5100209}}.
\newline\urlprefix\url{https://www.mdpi.com/2218-1997/5/10/209}

\bibitem{Hillery1999}
M.~Hillery, V.~Bu{\v{z}}ek, A.~Berthiaume,
  \href{https://link.aps.org/doi/10.1103/PhysRevA.59.1829}{{Quantum secret
  sharing}}, Physical Review A 59~(3) (1999) 1829--1834 (mar 1999).
\newblock \href {https://doi.org/10.1103/PhysRevA.59.1829}
  {\path{doi:10.1103/PhysRevA.59.1829}}.
\newline\urlprefix\url{https://link.aps.org/doi/10.1103/PhysRevA.59.1829}

\bibitem{PhysRevA.58.4394}
A.~Karlsson, M.~Bourennane,
  \href{https://link.aps.org/doi/10.1103/PhysRevA.58.4394}{Quantum
  teleportation using three-particle entanglement}, Phys. Rev. A 58 (1998)
  4394--4400 (Dec 1998).
\newblock \href {https://doi.org/10.1103/PhysRevA.58.4394}
  {\path{doi:10.1103/PhysRevA.58.4394}}.
\newline\urlprefix\url{https://link.aps.org/doi/10.1103/PhysRevA.58.4394}

\bibitem{Reed2012}
M.~D. Reed, L.~DiCarlo, S.~E. Nigg, L.~Sun, L.~Frunzio, S.~M. Girvin, R.~J.
  Schoelkopf, \href{https://doi.org/10.1038/nature10786}{Realization of
  three-qubit quantum error correction with superconducting circuits}, Nature
  482~(7385) (2012) 382--385 (2012).
\newblock \href {https://doi.org/10.1038/nature10786}
  {\path{doi:10.1038/nature10786}}.
\newline\urlprefix\url{https://doi.org/10.1038/nature10786}

\bibitem{Bose2003a}
S.~Bose, \href{http://link.aps.org/doi/10.1103/PhysRevLett.91.207901}{{Quantum
  Communication through an Unmodulated Spin Chain}}, Physical Review Letters
  91~(20) (2003) 1--4 (nov 2003).
\newblock \href {https://doi.org/10.1103/PhysRevLett.91.207901}
  {\path{doi:10.1103/PhysRevLett.91.207901}}.
\newline\urlprefix\url{http://link.aps.org/doi/10.1103/PhysRevLett.91.207901}

\bibitem{doi:10.1142/S0217979213450355}
T.~J.~G. APOLLARO, S.~LORENZO, F.~PLASTINA,
  \href{https://doi.org/10.1142/S0217979213450355}{Transport of quantum
  correlations across a spin chain}, International Journal of Modern Physics B
  27~(01n03) (2013) 1345035 (2013).
\newblock \href
  {http://arxiv.org/abs/https://doi.org/10.1142/S0217979213450355}
  {\path{arXiv:https://doi.org/10.1142/S0217979213450355}}, \href
  {https://doi.org/10.1142/S0217979213450355}
  {\path{doi:10.1142/S0217979213450355}}.
\newline\urlprefix\url{https://doi.org/10.1142/S0217979213450355}

\bibitem{Banchi2011a}
L.~Banchi, T.~J.~G. Apollaro, a.~Cuccoli, R.~Vaia, P.~Verrucchi,
  \href{http://stacks.iop.org/1367-2630/13/i=12/a=123006?key=crossref.57b551843a2ded2fda892c8412520260}{{Long
  quantum channels for high-quality entanglement transfer}}, New Journal of
  Physics 13~(12) (2011) 123006 (dec 2011).
\newblock \href {https://doi.org/10.1088/1367-2630/13/12/123006}
  {\path{doi:10.1088/1367-2630/13/12/123006}}.
\newline\urlprefix\url{http://stacks.iop.org/1367-2630/13/i=12/a=123006?key=crossref.57b551843a2ded2fda892c8412520260}

\bibitem{Apollaro2015}
T.~J.~G. Apollaro, S.~Lorenzo, A.~Sindona, S.~Paganelli, G.~L. Giorgi,
  F.~Plastina,
  \href{http://stacks.iop.org/1402-4896/2015/i=T165/a=014036?key=crossref.5171a929b28f4ca4dd1cf89861778fd8}{{Many-qubit
  quantum state transfer via spin chains}}, Physica Scripta T165~(T165) (2015)
  014036 (oct 2015).
\newblock \href {https://doi.org/10.1088/0031-8949/2015/T165/014036}
  {\path{doi:10.1088/0031-8949/2015/T165/014036}}.
\newline\urlprefix\url{http://stacks.iop.org/1402-4896/2015/i=T165/a=014036?key=crossref.5171a929b28f4ca4dd1cf89861778fd8}

\bibitem{Almeida2017}
G.~M.~A. Almeida, F.~A. B.~F. de~Moura, T.~J.~G. Apollaro, M.~L. Lyra,
  \href{http://arxiv.org/abs/1707.05865}{{Disorder-assisted distribution of
  entanglement in {\$}XY{\$} spin chains}} 032315 (2017) 1--10 (2017).
\newblock \href {http://arxiv.org/abs/1707.05865} {\path{arXiv:1707.05865}},
  \href {https://doi.org/10.1103/PhysRevA.96.032315}
  {\path{doi:10.1103/PhysRevA.96.032315}}.
\newline\urlprefix\url{http://arxiv.org/abs/1707.05865}

\bibitem{VIEIRA20182586}
R.~Vieira, G.~Rigolin,
  \href{http://www.sciencedirect.com/science/article/pii/S0375960118307710}{{Almost
  perfect transport of an entangled two-qubit state through a spin chain}},
  Physics Letters A 382~(36) (2018) 2586--2594 (2018).
\newblock \href
  {https://doi.org/https://doi.org/10.1016/j.physleta.2018.07.027}
  {\path{doi:https://doi.org/10.1016/j.physleta.2018.07.027}}.
\newline\urlprefix\url{http://www.sciencedirect.com/science/article/pii/S0375960118307710}

\bibitem{Vieira2019}
R.~Vieira, G.~Rigolin,
  \href{http://arxiv.org/abs/1903.11419{\%}0Ahttp://dx.doi.org/10.1007/s11128-019-2254-1}{{Robust
  and efficient transport of two-qubit entanglement via disordered spin
  chains}}, Quantum Information Processing 123 (2019).
\newblock \href {http://arxiv.org/abs/1903.11419} {\path{arXiv:1903.11419}},
  \href {https://doi.org/10.1007/s11128-019-2254-1}
  {\path{doi:10.1007/s11128-019-2254-1}}.
\newline\urlprefix\url{http://arxiv.org/abs/1903.11419{\%}0Ahttp://dx.doi.org/10.1007/s11128-019-2254-1}

\bibitem{chetcuti2019perturbativelyperfect}
W.~J. Chetcuti, C.~Sanavio, S.~Lorenzo, T.~J.~G. Apollaro,
  Perturbatively-perfect many-body transfer (2019).
\newblock \href {http://arxiv.org/abs/1911.12211} {\path{arXiv:1911.12211}}.

\bibitem{Wojcik2007}
A.~W{\'{o}}jcik, T.~{\L}uczak, P.~Kurzy{\'{n}}ski, A.~Grudka, T.~Gdala,
  M.~Bednarska,
  \href{http://link.aps.org/doi/10.1103/PhysRevA.75.022330}{{Multiuser quantum
  communication networks}}, Physical Review A 75~(2) (2007) 022330 (feb 2007).
\newblock \href {https://doi.org/10.1103/PhysRevA.75.022330}
  {\path{doi:10.1103/PhysRevA.75.022330}}.
\newline\urlprefix\url{http://link.aps.org/doi/10.1103/PhysRevA.75.022330}

\bibitem{Lorenzo2016}
S.~Lorenzo, T.~J.~G. Apollaro, A.~Trombettoni, S.~Paganelli, {Quantum state
  transfer with ultracold atoms in optical lattices} (2016) 1--14 (2016).
\newblock \href {http://arxiv.org/abs/1610.03248} {\path{arXiv:1610.03248}}.

\bibitem{qst2}
S.~Lorenzo, T.~Apollaro, S.~Paganelli, G.~Palma, F.~Plastina, {Transfer of
  arbitrary two qubit states via a spin chain}, Phys. Rev. A 91 (2015) 42321
  (2015).

\bibitem{Lieb1961}
E.~Lieb, T.~Schultz, D.~Mattis,
  \href{http://linkinghub.elsevier.com/retrieve/pii/0003491661901154}{{Two
  soluble models of an antiferromagnetic chain}}, Annals of Physics 16~(3)
  (1961) 407--466 (dec 1961).
\newblock \href {https://doi.org/10.1016/0003-4916(61)90115-4}
  {\path{doi:10.1016/0003-4916(61)90115-4}}.
\newline\urlprefix\url{http://linkinghub.elsevier.com/retrieve/pii/0003491661901154}

\bibitem{PhysRevLett.80.2245}
W.~K. Wootters,
  \href{https://link.aps.org/doi/10.1103/PhysRevLett.80.2245}{Entanglement of
  formation of an arbitrary state of two qubits}, Phys. Rev. Lett. 80 (1998)
  2245--2248 (Mar 1998).
\newblock \href {https://doi.org/10.1103/PhysRevLett.80.2245}
  {\path{doi:10.1103/PhysRevLett.80.2245}}.
\newline\urlprefix\url{https://link.aps.org/doi/10.1103/PhysRevLett.80.2245}

\bibitem{Acin2001a}
A.~Ac{\'{i}}n, D.~Bru{\ss}, M.~Lewenstein, A.~Sanpera,
  \href{https://link.aps.org/doi/10.1103/PhysRevLett.87.040401}{{Classification
  of Mixed Three-Qubit States}}, Physical Review Letters 87~(4) (2001) 040401
  (jul 2001).
\newblock \href {https://doi.org/10.1103/PhysRevLett.87.040401}
  {\path{doi:10.1103/PhysRevLett.87.040401}}.
\newline\urlprefix\url{https://link.aps.org/doi/10.1103/PhysRevLett.87.040401}

\bibitem{PhysRevA.61.052306}
V.~Coffman, J.~Kundu, W.~K. Wootters,
  \href{https://link.aps.org/doi/10.1103/PhysRevA.61.052306}{Distributed
  entanglement}, Phys. Rev. A 61 (2000) 052306 (Apr 2000).
\newblock \href {https://doi.org/10.1103/PhysRevA.61.052306}
  {\path{doi:10.1103/PhysRevA.61.052306}}.
\newline\urlprefix\url{https://link.aps.org/doi/10.1103/PhysRevA.61.052306}

\bibitem{PhysRevA.83.062325}
Z.-H. Ma, Z.-H. Chen, J.-L. Chen, C.~Spengler, A.~Gabriel, M.~Huber,
  \href{https://link.aps.org/doi/10.1103/PhysRevA.83.062325}{Measure of genuine
  multipartite entanglement with computable lower bounds}, Phys. Rev. A 83
  (2011) 062325 (Jun 2011).
\newblock \href {https://doi.org/10.1103/PhysRevA.83.062325}
  {\path{doi:10.1103/PhysRevA.83.062325}}.
\newline\urlprefix\url{https://link.aps.org/doi/10.1103/PhysRevA.83.062325}

\bibitem{doi:10.1063/1.1497700}
D.~A. Meyer, N.~R. Wallach, \href{https://doi.org/10.1063/1.1497700}{Global
  entanglement in multiparticle systems}, Journal of Mathematical Physics
  43~(9) (2002) 4273--4278 (2002).
\newblock \href {http://arxiv.org/abs/https://doi.org/10.1063/1.1497700}
  {\path{arXiv:https://doi.org/10.1063/1.1497700}}, \href
  {https://doi.org/10.1063/1.1497700} {\path{doi:10.1063/1.1497700}}.
\newline\urlprefix\url{https://doi.org/10.1063/1.1497700}

\bibitem{PhysRevLett.114.160501}
G.~T\'oth, T.~Moroder, O.~G\"uhne,
  \href{https://link.aps.org/doi/10.1103/PhysRevLett.114.160501}{Evaluating
  convex roof entanglement measures}, Phys. Rev. Lett. 114 (2015) 160501 (Apr
  2015).
\newblock \href {https://doi.org/10.1103/PhysRevLett.114.160501}
  {\path{doi:10.1103/PhysRevLett.114.160501}}.
\newline\urlprefix\url{https://link.aps.org/doi/10.1103/PhysRevLett.114.160501}

\bibitem{TERHAL2000319}
B.~M. Terhal,
  \href{http://www.sciencedirect.com/science/article/pii/S0375960100004011}{Bell
  inequalities and the separability criterion}, Physics Letters A 271~(5)
  (2000) 319 -- 326 (2000).
\newblock \href {https://doi.org/https://doi.org/10.1016/S0375-9601(00)00401-1}
  {\path{doi:https://doi.org/10.1016/S0375-9601(00)00401-1}}.
\newline\urlprefix\url{http://www.sciencedirect.com/science/article/pii/S0375960100004011}

\bibitem{PhysRevLett.92.087902}
M.~Bourennane, M.~Eibl, C.~Kurtsiefer, S.~Gaertner, H.~Weinfurter, O.~G\"uhne,
  P.~Hyllus, D.~Bru\ss{}, M.~Lewenstein, A.~Sanpera,
  \href{https://link.aps.org/doi/10.1103/PhysRevLett.92.087902}{Experimental
  detection of multipartite entanglement using witness operators}, Phys. Rev.
  Lett. 92 (2004) 087902 (Feb 2004).
\newblock \href {https://doi.org/10.1103/PhysRevLett.92.087902}
  {\path{doi:10.1103/PhysRevLett.92.087902}}.
\newline\urlprefix\url{https://link.aps.org/doi/10.1103/PhysRevLett.92.087902}

\bibitem{Jungnitsch2011}
B.~Jungnitsch, T.~Moroder, O.~G{\"{u}}hne,
  \href{http://link.aps.org/doi/10.1103/PhysRevLett.106.190502}{{Taming
  Multiparticle Entanglement}}, Physical Review Letters 106~(19) (2011) 1--4
  (may 2011).
\newblock \href {https://doi.org/10.1103/PhysRevLett.106.190502}
  {\path{doi:10.1103/PhysRevLett.106.190502}}.
\newline\urlprefix\url{http://link.aps.org/doi/10.1103/PhysRevLett.106.190502}

\bibitem{PhysRevA.99.012319}
B.~i. e. i. f. m.~c. \ifmmode~\mbox{\c{C}}\else \c{C}\fi{}akmak, S.~Campbell,
  B.~Vacchini, O.~E. M\"ustecapl\ifmmode \imath \else \i
  \fi{}o\ifmmode~\breve{g}\else \u{g}\fi{}lu, M.~Paternostro,
  \href{https://link.aps.org/doi/10.1103/PhysRevA.99.012319}{Robust
  multipartite entanglement generation via a collision model}, Phys. Rev. A 99
  (2019) 012319 (Jan 2019).
\newblock \href {https://doi.org/10.1103/PhysRevA.99.012319}
  {\path{doi:10.1103/PhysRevA.99.012319}}.
\newline\urlprefix\url{https://link.aps.org/doi/10.1103/PhysRevA.99.012319}

\bibitem{Eltschka_2014}
C.~Eltschka, J.~Siewert,
  \href{https://doi.org/10.1088%2F1751-8113%2F47%2F42%2F424005}{Quantifying
  entanglement resources}, Journal of Physics A: Mathematical and Theoretical
  47~(42) (2014) 424005 (oct 2014).
\newblock \href {https://doi.org/10.1088/1751-8113/47/42/424005}
  {\path{doi:10.1088/1751-8113/47/42/424005}}.
\newline\urlprefix\url{https://doi.org/10.1088%2F1751-8113%2F47%2F42%2F424005}

\bibitem{YU2004377}
C.~shui Yu, H.~shan Song,
  \href{http://www.sciencedirect.com/science/article/pii/S0375960104010722}{Free
  entanglement measure of multiparticle quantum states}, Physics Letters A
  330~(5) (2004) 377 -- 383 (2004).
\newblock \href
  {https://doi.org/https://doi.org/10.1016/j.physleta.2004.07.054}
  {\path{doi:https://doi.org/10.1016/j.physleta.2004.07.054}}.
\newline\urlprefix\url{http://www.sciencedirect.com/science/article/pii/S0375960104010722}

\bibitem{Sabin2008}
C.~Sab{\'{i}}n, G.~Garc{\'{i}}a-Alcaine,
  \href{http://www.springerlink.com/index/10.1140/epjd/e2008-00112-5}{{A
  classification of entanglement in three-qubit systems}}, The European
  Physical Journal D 48~(3) (2008) 435--442 (jun 2008).
\newblock \href {https://doi.org/10.1140/epjd/e2008-00112-5}
  {\path{doi:10.1140/epjd/e2008-00112-5}}.
\newline\urlprefix\url{http://www.springerlink.com/index/10.1140/epjd/e2008-00112-5}

\bibitem{Vidal2002a}
G.~Vidal, R.~F. Werner, {Computable measure of entanglement}, Physical Review A
  - Atomic, Molecular, and Optical Physics 65~(3) (2002) 1--11 (2002).
\newblock \href {http://arxiv.org/abs/0102117} {\path{arXiv:0102117}}, \href
  {https://doi.org/10.1103/PhysRevA.65.032314}
  {\path{doi:10.1103/PhysRevA.65.032314}}.

\bibitem{HORODECKI19961}
M.~Horodecki, P.~Horodecki, R.~Horodecki,
  \href{http://www.sciencedirect.com/science/article/pii/S0375960196007062}{Separability
  of mixed states: necessary and sufficient conditions}, Physics Letters A
  223~(1) (1996) 1 -- 8 (1996).
\newblock \href {https://doi.org/https://doi.org/10.1016/S0375-9601(96)00706-2}
  {\path{doi:https://doi.org/10.1016/S0375-9601(96)00706-2}}.
\newline\urlprefix\url{http://www.sciencedirect.com/science/article/pii/S0375960196007062}

\bibitem{Dur1999}
W.~D{\"{u}}r, J.~I. Cirac, R.~Tarrach, {Separability and distillability of
  multiparticle quantum systems}, Physical Review Letters 83~(17) (1999)
  3562--3565 (1999).
\newblock \href {http://arxiv.org/abs/9903018} {\path{arXiv:9903018}}, \href
  {https://doi.org/10.1103/PhysRevLett.83.3562}
  {\path{doi:10.1103/PhysRevLett.83.3562}}.

\bibitem{Amico2004b}
L.~Amico, A.~Osterloh, F.~Plastina, R.~Fazio, G.~{Massimo Palma},
  \href{http://link.aps.org/doi/10.1103/PhysRevA.69.022304}{{Dynamics of
  entanglement in one-dimensional spin systems}}, Physical Review A 69~(2)
  (2004) 022304 (feb 2004).
\newblock \href {https://doi.org/10.1103/PhysRevA.69.022304}
  {\path{doi:10.1103/PhysRevA.69.022304}}.
\newline\urlprefix\url{http://link.aps.org/doi/10.1103/PhysRevA.69.022304}

\bibitem{DBLP:journals/qic/LaustsenVE03}
T.~Laustsen, F.~Verstraete, S.~J. van Enk,
  \href{http://portal.acm.org/citation.cfm?id=2011514}{Local vs. joint
  measurements for the entanglement of assistance}, Quantum Information {\&}
  Computation 3~(1) (2003) 64--83 (2003).
\newline\urlprefix\url{http://portal.acm.org/citation.cfm?id=2011514}

\end{thebibliography}

\end{document}